\documentclass{article}
\usepackage{amssymb,slashed,framed}

\usepackage{graphicx}
\usepackage{amsmath,mathrsfs}

\newtheorem{theorem}{Theorem}

\newtheorem{proposition}[theorem]{Proposition}
\newtheorem{remark}[theorem]{Remark}

\begin{document}

\title{Quantum Time Travel Revisited:\\
Noncommutative M\"{o}bius Transformations\\ and Time Loops}

\author{John E. Gough\\
 Department of Physics, Aberystwyth University, \\
 Wales, UK, SY23 3QR.}

\date{https://orcid.org/0000-0002-1374-328X}

\maketitle

\begin{abstract}
We extend the theory of quantum time loops introduced by Greenberger and Svozil \cite{GS05} from the scalar situation (where paths have just an associated complex amplitude) to the general situation where the time traveling system has multi-dimensional underlying Hilbert space. The main mathematical tool which emerges is the noncommutative M\"{o}bius Transformation and this affords a formalism similar to the modular structure well known to feedback control problems. The self-consistency issues that plague other approaches do not arise in this approach as we do not consider completely closed time loops.
We argue that a sum-over-all-paths approach may be carried out in the scalar case, but quickly becomes unwieldy in the general case. It is natural to replace the beamsplitters of \cite{GS05} with more general components having their own quantum structure, in which case the theory starts to resemble the quantum feedback networks theory for open quantum optical models and indeed we exploit this to look at more realistic physical models of time loops. We analyze some Grandfather paradoxes in the new setting.
\end{abstract}


\section{Introduction}
Closed timelike curves are known to exist for several solution to the equations of General Relativity and have been the focus of considerable interest \cite{Weak_Energy,Novikov89,Frolov_Novikov,F90,Echeverria}. While some authors are more forthcoming about this than others, the topic is really about time machines: as Hawking remarks, ``even if it turns out that time travel is impossible, it is important that we understand why it is impossible'' \cite{Future}. The formulation of the laws of thermodynamics in terms of the impossibility of perpetual motion machines being a motivation. 

The question about the quantum formulation of the arrow of time has being of long standing importance \cite{Deutsch,Schulman,Lloyd11,t'Hooft}. The specific issue of time travel paradoxes and quantum mechanics was addressed by Pegg \cite{Pegg01}, and the prospect of quantum systems undergoing time loops was raised Greenberger and Svozil \cite{GS05}. We shall generalize their analysis by assuming that the dynamic variables of a time-traveling quantum particle (its position and momentum, etc.) can be decoupled from the internal degrees of freedom (for instance, spin, polarization, etc.). 

Indeed, the paths of the quantum particle may be considered ``classical'' in the sense that our resolution of distances is well above whatever de Broglie wavelength may be assigned to the particle, see for instance \cite{Frigerio_Ruzzier}. Indeed, we could view the paths themselves as some form of fiber cable which constraints the motion of the quantum system (we make no attempt to explain the backward-in-time propagation!). Instead, the quantumness of the particle that is of interest here is not its motional behaviour but its internal structure: we take the underlying Hilbert space for the internal degrees of freedom to be $\mathfrak{h}$ and the system carries this quantum information along its travels. The quantum time machine technology is then manifest in the nodes where the paths meet and here we have generalized beamsplitter action on incoming (internal) quantum states. In \cite{GS05}, only phases are considered ($ \mathfrak{h} \cong \mathbb{C}$), however, we shall give the generalization to arbitrary separable Hilbert spaces.

We shall review this in section \ref{sec:QTL} and introduce the non-commutative M\"{o}bius transformation as the main mathematical device for handling general Hilbert spaces. Here, we are inspired by the theory of quantum feedback networks (sometimes referred to as the $SLH$ formalism) introduced in \cite{SLH,Series_Product}: for an overview see \cite{CKS}.

In section \ref{sec:Paths} we shall look at the scalar case ($ \mathfrak{h} \cong \mathbb{C}$) which is essentially the abelian framework. We shall use a \textit{welcher weg} path analysis, however, we quickly see that the M\"{o}bius transform provides a more efficient description. This may be somewhat cautionary given the prevalence of path integral methods for propagating quantum particles in spacetime.

In Section \ref{sec:Grandfather} we shall study various grandfather paradoxes (going back in time and killing your grandfather) in our language.

\section{Quantum Time Loops}
\label{sec:QTL}
\begin{figure}[ht]
\centering
\includegraphics[width=0.5\textwidth]{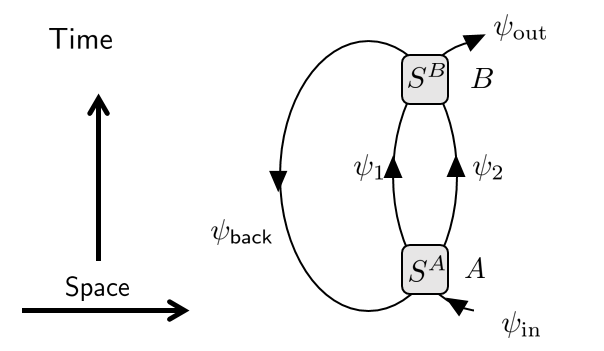}
\caption{A particle interacts with a future version of itself using two beamsplitters and time travel.}
\label{fig:GS_loop}
\end{figure}

In the set up of Greenberger and Svozil \cite{GS05}, a particle with probability amplitude $\psi _{\text{in}}$ enters into one port of a beamsplitter at spacetime event $A$. Another signal $\psi _{\text{back}}\left( t_{A}\right) $ enters at the other port.
They recombine at a beasmsplitter (possibly the same one!) at a later spacetime event $B$. One of the outputs, $\psi _{\text{back}}\left( t_{B}\right) $ is sent back in time and fed in as the original second initial
input at $A$. (See Figure \ref{fig:GS_loop}.)

The two beamsplitters effectively act as a Mach-Zehnder interferometer, now distributed over spacetime, with probability amplitudes $\psi_1$ and $\psi_2$ in the two branches of the interferometer. Note that we must have
\begin{eqnarray}
    | \psi_{\text{in} } (t_A )|^2=   | \psi_{\text{out} } (t_B )|^2
    \label{eq:phase}
\end{eqnarray}

As we also have space dimensions and so it is meaningful to say that the particle comes in from the right ($\psi_{\mathrm{in}}$) and goes out to the right ($\psi_{\mathrm{out}}$). Likewise, the time-traveling particle comes from the left at $A$ and exits to the left from $B$. The input-output relations are 
\begin{eqnarray}
\left[ 
\begin{array}{c}
\psi _{1}  \\ 
\psi _{2} 
\end{array}
\right] _A =S^A \left[ 
\begin{array}{c}
\psi _{\text{back}}  \\ 
\psi _{\text{in}} 
\end{array}
\right]_A ,
\left[ 
\begin{array}{c}
 \psi _{\text{back}} \\
\psi _{\text{out}} 
\end{array}
\right]_B =S^B \left[ 
\begin{array}{c}
\psi _{1} \\ 
\psi _{2}
\end{array}
\right]_B 
\end{eqnarray}
where $S^A$ and $S^B$ are the (unitary!) beamsplitter matrices at $A$ and $B$, respectively. Likewise we have the propagation
\begin{eqnarray}
\psi _{k}\left( t_{B}\right) &=&G_{k}\, \psi _{k}\left( t_{A}\right) ,\quad
\left( k=1,2\right) , \\
\psi _{\text{back}}\left( t_{A}\right) &=&M\, \psi _{\text{back}}\left(
t_{B}\right) ,
\end{eqnarray}
where the $G_{k}$ are the phases picked up along the paths of $\psi _{1}$ and $\psi _{2}$ from time $t_{A}$ to $t_{B}$, and $M$ is the phase picked along the backward path of $\psi _{\text{back}}$. We therefore have 
\begin{eqnarray}
\left[ 
\begin{array}{c}
\psi _{\text{back}}  \\ 
\psi _{\text{out}} 
\end{array}
\right]_B =S \, \left[ 
\begin{array}{c}
\psi _{\text{back}}  \\ 
\psi _{\text{in}} 
\end{array}
\right]_A
\label{eq:pre-link}
\end{eqnarray}
where 
\begin{eqnarray}
S=\left[ 
\begin{array}{cc}
S_{\text{back,back}} & S_{\text{back},\text{in}} \\ 
S_{\text{out,}\text{back}} & S_{\text{out,in}}
\end{array}
\right] = S^B 
\left[ 
\begin{array}{cc}
G_1 & 0 \\ 
0 & G_2
\end{array}
\right] 
S^A .
\label{eq:S_tot}
\end{eqnarray}
Note that $S_{jk}=\sum_{i=1,2}S_{ji}^{B}G_i S_{ik}^{A}$.

They eliminate the backward path amplitudes to arrive at the overall input-output relation
\begin{eqnarray}
\psi _{\text{out}}\left( t_{B}\right) =S_{\text{fb}}\,\psi _{\text{in}%
}\left( t_{A}\right)
\end{eqnarray}
where the feedback ``gain'' is 
\begin{eqnarray}
S_{\text{fb}} =  
S_{\text{out,in}}+
\frac{ S_{\text{out,back} } M S_{\text{back,in} } }
{1-S_{\text{back,back}} M}.
\label{eq:S_fb}
\end{eqnarray}
The time loop is well-posed so long as $1-S_{\text{back,back}} M$ is invertible, in which case $S_{\text{fb}}$ works out to be a unimodular phase, thereby ensuring (\ref{eq:phase}). 

We shall now show that this generalizes to particles with internal degrees of freedom, that is, described by states $\vert \psi \rangle$ in a fixed Hilbert space, $\mathfrak{h}$.


\subsection{M\"{o}bius Transformations in Hilbert Space}
\label{sec:mobius}
Our main mathematical tool shall be the (noncommutative) M\"{o}bius transformation. Suppose that we have a pair of Hilbert spaces $\mathfrak{h}_{1}$ and $\mathfrak{h}_{2}$. Let us form their direct sum $\mathfrak{h}_{1}\oplus \mathfrak{h}_{2}$ and here we will use the standard representation
\begin{eqnarray}
|\phi _{1}\rangle \oplus |\phi _{2}\rangle \equiv \left[ 
\begin{array}{c}
|\phi _{1}\rangle  \\ 
|\phi _{2}\rangle 
\end{array}
\right] .
\end{eqnarray}
Within this representation, any linear operator on $\mathfrak{h}_{1}\oplus \mathfrak{h}_{2}$ may be decomposed as 
\begin{eqnarray}
S \equiv \left[ 
\begin{array}{cc}
S_{11} & S_{12} \\ 
S_{21} & S_{22}
\end{array}
\right] 
\end{eqnarray}
where the block $S_{ij}$ is a linear operator from $\mathfrak{h}_{j}$ to $\mathfrak{h}_{i}$. We say the operator is nontrivial if it is not block diagonal, that is, $S_{12}\neq 0$ and $S_{21}\neq 0$. 
In detail, $|\phi _{1}^\prime \rangle \oplus |\phi _{2}^\prime \rangle = S \, |\phi _{1}\rangle \oplus |\phi _{2}\rangle$ with \begin{eqnarray*}
    |\phi _{1}^\prime \rangle &=&   S_{11} |\phi _{1}\rangle + S_{12} |\phi _2 \rangle , \\
    |\phi _{2}^\prime \rangle &=&   S_{21} |\phi _{1}\rangle + S_{22} |\phi _{2} \rangle .
\end{eqnarray*}

Now let us impose the constraint $| \phi_2 \rangle = X | \phi_2^\prime \rangle$ for $X$ a fixed linear operator on $\mathfrak{h}_2$, then we may eliminate $| \phi_2^\prime \rangle$ to obtain $ | \phi_2 \rangle = (1-XS_{22} )^{-1} | \phi_1 \rangle$, provided the inverse exists. The result is that 
\begin{eqnarray}
    | \phi_1^\prime \rangle = \text{M\"{o}b} _1 ( S , X) \, | \phi_1 \rangle ,
\end{eqnarray}
where the \textit{noncommutative M\"{o}bius transformation} is the operator on $\mathfrak{h}_1$ defined by
\begin{eqnarray}
\text{M\"{o}b} _1 ( S , X) = S_{11} +S_{12}  (1- XS_{22}  )^{-1} X S_{21} .
\label{eq:Mobius}
\end{eqnarray}
Specifically, $\text{M\"{o}b} _1$ means that we have ``short-circuited'' Hilbert space $\mathfrak{h}_2$ and retain an operator on $\mathfrak{h}_1$ only. We will drop the subscript when it is obvious which Hilbert space is being shorted. It should be emphasized that none of the blocks are assume to commute! Nevertheless, whenever it exist (we shall say well-posed), it has some striking properties. 

\bigskip

\textbf{Siegel's Lemma}
If $S$ is a unitary on $\mathfrak{h}_1 \oplus \mathfrak{h}_2$ and $X$ is unitary on $\mathfrak{h}_2$, then provided the M\"{o}bius transformation is well-posed $\text{M\"{o}b} _1 ( S , X)$ is a unitary on $\mathfrak{h}_1$.

\bigskip

This is an old result due to C.L. Siegel and a proof can be found, for instance, in Young \cite{Young}. If either $S$ or $X$ are replaced with contractions, then $\text{M\"{o}b} _1 ( S , X)$
will likewise be a contraction. 
Alternative forms, assuming the inverses exists, are
\begin{eqnarray}
\text{M\"{o}b} _1 ( S , X) &=& S_{11} +S_{12} X (1- S_{22} X )^{-1} S_{21} \\
&=& S_{11} +S_{12}  ( X^{-1} - S_{22} )^{-1} S_{21}  
,
\end{eqnarray}
along with the \textit{multi-loop} expansion
\begin{eqnarray}
\text{M\"{o}b} _1 ( S , X) = S_{11} + \sum_{n=0}^\infty S_{12} X \big( S_{22} X \big)^n S_{21}  .
\label{eq:multiloop_expansion}
\end{eqnarray}

\subsection{Simple Time Loops}
The set up of Greenberger and Svozil \cite{GS05} deals with complex amplitudes and the phase $S_{\text{fb}}$ they calculate is a commutative M\"{o}bius transformation, however, it is not difficult to see that this generalizes to the case where the input particle has state $\vert \phi_{\mathrm{in}} \rangle $ belonging to a Hilbert space $\mathfrak{h}_{\mathrm{ext}}$ which we call the \textit{external Hilbert space}. We shall also have an \textit{internal Hilbert space} $\mathfrak{h}_{\mathrm{int}}$ for the time traveling particle, and this will be isomorphic to $\mathfrak{h}_{\mathrm{int}}$ in this situation.

\begin{figure}[ht]
	\centering
		\includegraphics[width=0.50\textwidth]{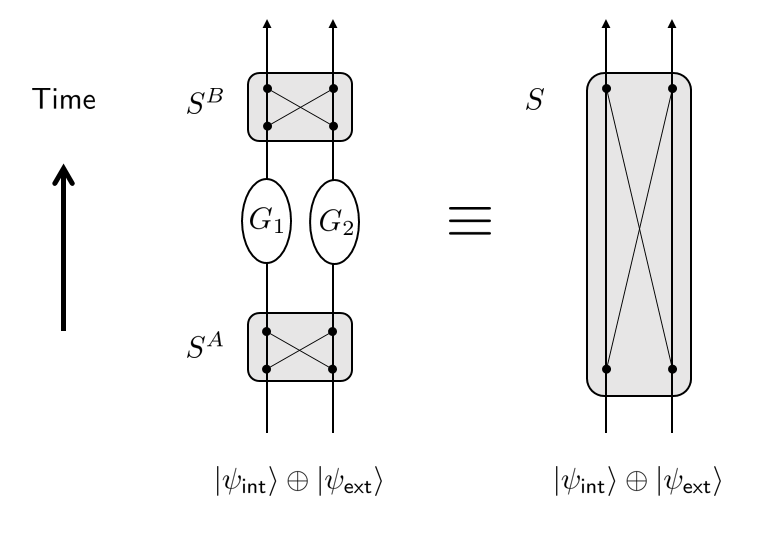}
	\caption{The open loop unitary operator $S$.}
	\label{fig:Series}
\end{figure}

Our first step will be to compute the \textit{open loop} unitary $S\in U\left( \mathfrak{h}_{\text{int}}\oplus \mathfrak{h}_{\text{ext}}\right) $ describing the evolution of the particles from time $t_A$ to time $t_B$, see Figure \ref{fig:Series}. This time, we make no assumptions about the commutativity of the blocks $S^A{ij}, G_k, S^B_{ij}$. The total forward evolution is then given by the unitary
\begin{eqnarray*}
S=S^{B}\left[ 
\begin{array}{cc}
G_{1} & 0 \\ 
0 & G_{2}
\end{array}
\right] S^{A}=\left[ 
\begin{array}{cc}
S_{\text{ext,ext}} & S_{\text{ext,int}} \\ 
S_{\text{int,ext}} & S_{\text{int,int}}
\end{array}
\right] 
\end{eqnarray*}
where $S^{A}$ and $S^{B}$ are the elements at the events $A$ and $B$ respectively. Overall, $S$ should yield another unitary. Schematically, we may describe this as in Figure \ref{fig:Series}.

\begin{figure}[htb]
	\centering
		\includegraphics[width=0.50\textwidth]{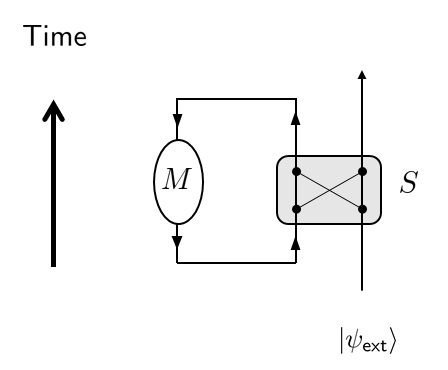}
	\caption{The time loop as a feedback system.}
	\label{fig:FLT}
\end{figure}

The output $S|\psi _{\text{int}}\rangle \oplus |\psi _{\text{ext}}\rangle $ can then be written as $|\psi _{\text{int}}^{\prime }\rangle \oplus |\psi _{\text{ext}}^{\prime }\rangle $ and the time loop is established by taking $|\psi _{\text{int}}\rangle \equiv M|\psi _{\text{int}}^{\prime }\rangle $, see Figure \ref{fig:FLT}. 

Again, eliminating the internal state leads to $|\psi _{\text{ext}}^{\prime }\rangle =S_{\text{fb}}|\psi _{\text{ext}}\rangle $ where 
\begin{eqnarray}
S_{\text{fb}} =
\text{M\"{o}b}_{\mathrm{ext}} ( S, M) .
\end{eqnarray}
This is the generalization of the commutative formula (\ref{eq:S_fb}).

\section{Comparison with Previous Theories}
As mentioned, our approach generalizes that of Greenberger and Svozil \cite{GS05} and it is worth emphasizing the mathematical structure here. Our time loops are generally feedback loops describing incoming and outgoing systems (inputs and outputs). 

\begin{figure}[ht]
	\centering
		\includegraphics[width=0.750\textwidth]{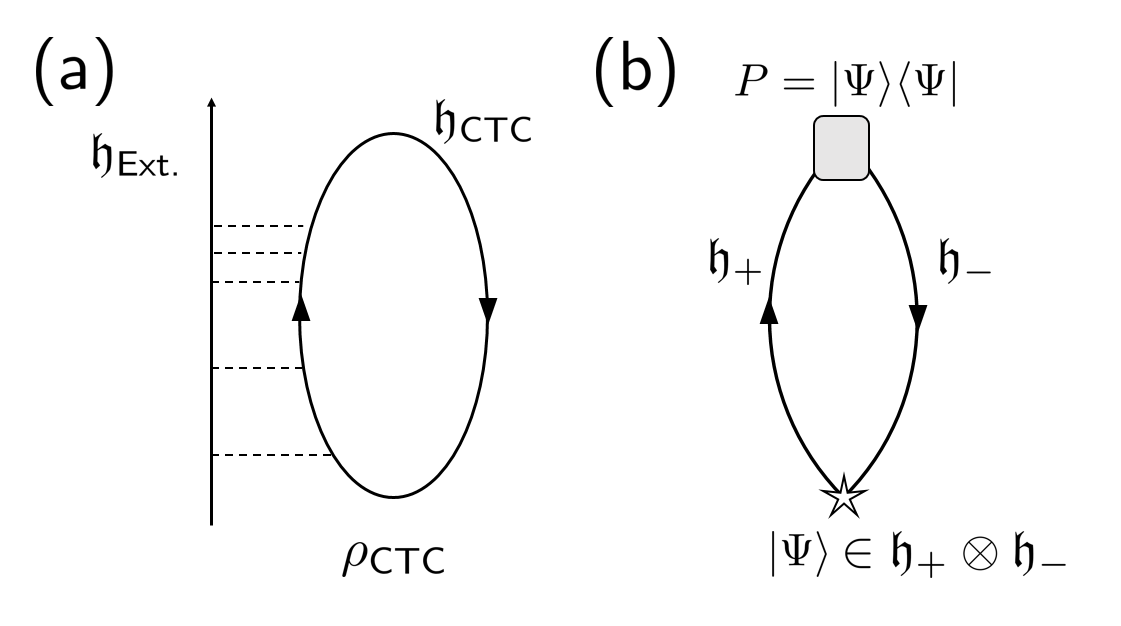}
	\caption{Completely Closed Timelike Curves. (a) Deutsch closed timelike curve (CTC) with underlying Hilbert space $\mathfrak{h}_{\text{CTC}}$ It interacts with its external environment with Hilbert space $\mathfrak{h}_{\text{Ext.}}$ represented by several lines leading to a total unitary evolution $U$ on the joint Hilbert space. (b) The time loop is formed when a particle-antiparticle pair come into existence at the event labeled as a star. They then annihilate later. The antiparticle is interpreted as a copy of the original system traveling back in time.}
	\label{fig:CTC_loops}
\end{figure}

This contrasts with the constructions given elsewhere which do not have external inputs or outputs, see Figure \ref{fig:CTC_loops}, and we outline these below.

\subsection{Deutschian Time-Loops}
In the Theory of Deutsch \cite{Deutsch}, the time traveling component comprises an isolated loop system with no input or output state in our sense. Instead, it is a quantum system with an underlying Hilbert space $\mathfrak{h}_{\text{CTC}}$ traveling around a closed timelike curve, see Figure \ref{fig:CTC_loops} (a). In addition to the loop, there will be an external environment (assumed to be chronology respecting) with Hilbert space $\mathfrak{h}_{\mathrm{Ext.}}$ The time traveling particle (on the future oriented part of its journey) may be assume to interact with the external environment via unitary $U$ acting on the tensor product space $ \mathfrak{h}_{\mathrm{Ext.}} \otimes \mathfrak{h}_{\text{CTC}}$. For a fixed external state $\rho_{\text{Ext.}}$, Deutsch imposes the self-consistency condition on the state of the time traveling system's state $\rho_{\text{CTC}}$:
\begin{eqnarray}
    \rho_{\text{CTC}} = \text{tr}_{\mathrm{Ext.}}
    \bigg\{ U ( \rho_{\text{Ext.}} \otimes \rho_{\text{CTC}}  ) U^\ast \bigg\} .
\end{eqnarray}
This ensures that internal state of the time traveler returns to its ``initial state'', and this is interpreted as enforcing the Novikov self-consistency condition \cite{Novikov89,Frolov_Novikov,F90}. 

The state $\rho_{\text{CTC}}$ is typically mixed and there is no mechanism proposed as to how it physical emerges.

\subsection{Lloydian Time Loops}
In the Theory of Lloyd \textit{et alia} \cite{Lloyd11} we also encounter a closed timelike curve without inputs and outputs. This time, the underlying Hilbert space is decomposed as
\begin{eqnarray}
    \mathfrak{h}_{\text{CTC}} =\mathfrak{h}_+ \otimes \mathfrak{h}_- ,
\end{eqnarray}
 where $\mathfrak{h}_+$ describes the particle on the forward branch of the loop while, $\mathfrak{h}_-$ describes the particle on the backward branch. We have $\mathfrak{h}_+ \cong \mathfrak{h}_-$ and an appropriate way to think of this scheme is that the forward and backward branches actually describe a particle-antiparticle pair, see Figure \ref{fig:CTC_loops} (b). They propose a two-part self-consistency condition that does not dependent on the external environment: the first part assumes that when the pair come into existence at time $t_A$, then they will be in a (pure) maximally entangled state, say $| \Psi \rangle = \frac{1}{\sqrt{d}} \sum_{k=1}^d |e _k \rangle_+ \otimes |e_k \rangle _-$ where we have copies $\{ | e_k \rangle_\pm :k =1, \cdots , d\}$ of an orthonormal basis on $\mathfrak{h}_\pm$; the second part assumes that we perform a measurement at time $t_B$ which has the above state $|\Psi \rangle$ as eigenstate and we post-select this state.

\subsection{Our Time Loops}
In our situation, the loops are not isolated: they have an earlier version of the system which enters as input and a later version system that emerges as output. In principle, we may be using a device such as a wormhole, but our curves are not closed.

For different branches of the loop, we consider the \textit{direct sum} of their Hilbert spaces. This is the natural thing to do as the transfer from one branch to another is not an interaction between separate systems but instead the action of an exchange or scattering. We therefore do not have the self-consistency issues encountered in other theories. Instead, our only consistency condition is the well-posedness of the network, that is, the existence of the M\"{o}bius transformation (\ref{eq:Mobius}) however this requirement is standard for feedback control systems.

\section{Sum over Paths}
\label{sec:Paths}
\begin{figure}[ht]
	\centering
		\includegraphics[width=0.750\textwidth]{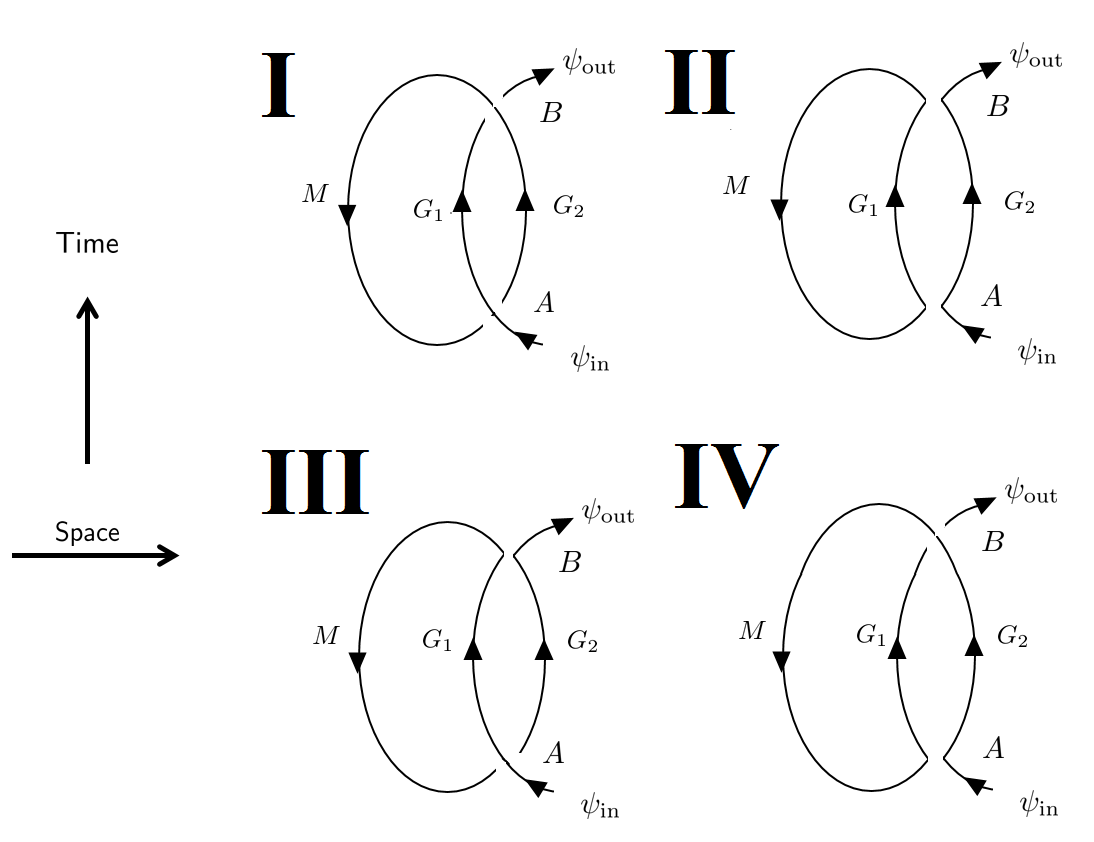}
	\caption{Path  I comes about from two reflections at both $A$ and $B$. In path II it is all transmissions. Paths III has transmissions at $A$ and reflections at $B$.}
	\label{fig:GS_paths}
\end{figure}

Let us reconsider the possible ``classical paths'' that a particle (or, as we shall see, two particles) may follow in the set up in Figure \ref{fig:GS_loop}. 
Depending on whether inputs are transmitted or reflected at each beamsplitter, we have four possible ``classical histories'', see Figure \ref{fig:GS_paths}, where no more than one particle is allowed down any leg of the path. 

Paths I and II each involve a pair of non-interacting particles, one of which is a closed curve. Paths III and IV involve a single particle which propagates forward, then backward, and again forward in time.

\begin{figure}[ht]
	\centering
		\includegraphics[width=0.75\textwidth]{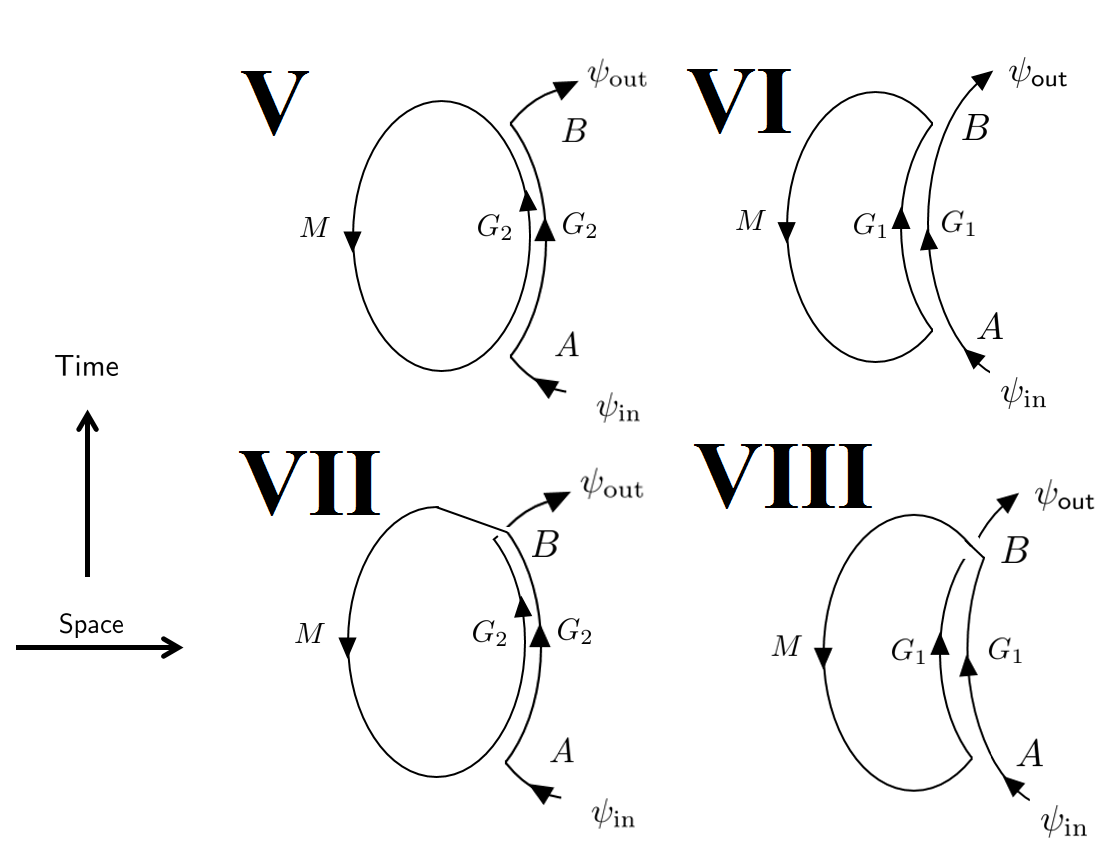}
	\caption{We obtain four more paths if we allow both particles to go down the same branch.}
	\label{fig:GS_paths2}
\end{figure}

There are four other paths which will contribute and these are the ones where both particles entering at $A$ go down the same branch, see Figure \ref{fig:GS_paths2}.

Physically, there will be no more contributions since, in order to have out incoming particle enter from the right and ultimately exit to the right, we need an even number of transmissions and of reflections.

We  remark that paths I, II, V and VI involved the particle traveling normally through time, though with some other particle caught in a time loop. Paths III, IV, VII and VIII necessitate the particle does some time traveling.

We can now assign transfer coefficients for each of the eight paths considered in Figures \ref{fig:GS_paths} and \ref{fig:GS_paths2}.

For instance, in path I we technically have two separate paths: the open-ended path has an (right-incoming) transmission, then a propagation along the first branch from $A$ to $B$, and another (right-incoming) transmission, yielding the \textit{feed  through transfer coefficient}
\begin{eqnarray}
G_{\mathrm{I}} = S_{\mathrm{out} , 1}^{B}G_1 S_{1, \mathrm{in}}^{A}  \equiv \kappa_{\mathrm{I}} \, G_1 .
\end{eqnarray}
The complex constants $\kappa_{\mathrm{I}}$ is a constant given as the product of the appropriate beamsplitter matrix elements for $A$ and $B$. 

We also have the closed curve and we associate the \textit{loop transfer coefficient} 
\begin{eqnarray}
C_{\mathrm{I}} = M S_{\mathrm{back} , 2}^{B}G_2 S_{2, \mathrm{back}}^{A}  \equiv \beta_{\mathrm{I}} MG_1
\end{eqnarray}
which comes from two (left-incoming) transmissions, and propagation forward in time along the first branch from $A$ to $B$, followed by a propagation back in time. Again, the constant $\beta_{\mathrm{I}}$ is the product of the two relevant beamsplitter matrix elements.

The full set of transfer coefficients is tabulated in Table \ref{tab:reg} below.

\begin{table}[ht]
\centering
\begin{tabular}{r|ll}
\textrm{Path}&
\textrm{Feed through}&
\textrm{Loop}\\
\hline
 	I &			$G_{\mathrm{I}} = \kappa_{\mathrm{I}} \, G_2$, & $C_{\mathrm{I}} = \beta_{\mathrm{I}} \, MG_1 $, \\
		  II &		$	G_{\mathrm{II}} = \kappa_{\mathrm{II}} \, G_1$, & $C_{\mathrm{II}} = \beta_{\mathrm{II}} \, MG_2 $, \\
			III & $G_{\mathrm{III}} = \kappa_{\mathrm{III}} \, G_2 M G_1 $, & $\,$ \\
			IV &		$	G_{\mathrm{IV}} = \kappa_{\mathrm{IV}} \, G_1 M G_2 $, & $\,$ \\
			V & $G_{\mathrm{V}} = G_{\mathrm{I}}$, & $C_{\mathrm{V}} =  \beta_{\mathrm{V}} \, MG_2 $, \\
			VI & $G_{\mathrm{VI}} = G_{\mathrm{II}}$, & $C_{\mathrm{VI}} = \beta_{\mathrm{VI}} \, MG_1$  ,\\
			VII & $G_{\mathrm{VII}} = \kappa_{\mathrm{VII}} \, G_2 M G_2 $, & $\,$ \\
			VIII & $G_{\mathrm{VII}} = \kappa_{\mathrm{VIII}} \,  G_1 M G_1 $, & 	$\,$
\end{tabular}
\caption{The path transfer coefficients.}
\label{tab:reg}
\end{table}

Let us note that 
\begin{eqnarray}
S_{\mathrm{out}, \mathrm{in} } = \sum_{i=1,2} S_{\mathrm{out} , i}^{B}G_{i} S_{i, \mathrm{in}}^{A} \equiv G_{\mathrm{I}} + G_{\mathrm{II}} ,
\end{eqnarray}
with $G_{\mathrm{I}}$ being the $i=1$ term, as seen above, and $G_{\mathrm{II}}$ the $i=2$ term.
Likewise,
\begin{eqnarray}
S_{\mathrm{out}, \mathrm{back}} M S_{\mathrm{back}, \mathrm{in}} \nonumber &=& \sum_{i=1,2} \sum_{j=1,2}
 S_{\mathrm{out} , i}^{B}G_i S_{i, \mathrm{back}}^{A} M S_{\mathrm{back} , j}^{B}G_j S_{j, \mathrm{in}}^{A} \nonumber \\
&&\equiv  G_{\mathrm{III}} + G_{\mathrm{IV}} + G_{\mathrm{VII}} +G_{\mathrm{VIII}} .
\end{eqnarray}
Here $G_{\mathrm{III}} , G_{\mathrm{IV}} , G_{\mathrm{VII}}$ and $G_{\mathrm{VIII}}$ arise as the terms $(i,j) = (2,1),(1,2), (2,2)$ and $(1,1)$, respectively.
Finally, we note that
\begin{eqnarray}
S_{\mathrm{back}, \mathrm{back}} M = \sum_{i=1,2}  
 S_{\mathrm{back} , i}^{B}G_{i} S_{i, \mathrm{back}}^{A} M  \equiv   C_{\mathrm{I}}  + C_{\mathrm{II}} .
\end{eqnarray}

We may put this together in the following statement.
\begin{proposition}
The transfer coefficient (\ref{eq:TF_GS}) then takes the form
\begin{eqnarray}
S_{\mathrm{fb}} = G_{\mathrm{I}} + G_{\mathrm{II}} + \frac{G_{\mathrm{III}}+G_{\mathrm{IV}}+G_{\mathrm{VII}}+G_{\mathrm{VIII}} } {1 - ( C_{\mathrm{I}} + C_{\mathrm{II}}) } .
\label{eq:S_frac}
\end{eqnarray}
\end{proposition}

The V and VI paths are conspicuous by their absence, however, we note that they have the same feed through transfers as I and II, respectively, and their loop transfers are proportional.

If we want to specify the propagators then this may be rewritten as
\begin{gather}
S_{\mathrm{fb}} = \kappa_{\mathrm{I}} \, G_1 + \kappa_{\mathrm{II}} \, G_2 
+ 
\frac{ \big( (\kappa_{\mathrm{III}} +\kappa_{\mathrm{IV}} )G_1G_2 +\kappa_{\mathrm{VII}}G_2^2+ \kappa_{\mathrm{VIII}} G_1^2 \big) M} {1 - ( \beta_{\mathrm{I}} G_1+ \beta_{\mathrm{II}}G_2) M} .
\label{eq:S_frac_prop}
\end{gather}

The formula (\ref{eq:S_frac_prop}) is, in and of itself, useful. It is linear in the feed through coefficients and fractional linear in the loop coefficients. We may expand the geometric series as $\sum_n (C_{\mathrm{I}} + C_{\mathrm{II}} )^n$, and then expand using the binomial theorem, so as to come up with a sum over all paths including multiple loop passes with various combinatorial factors (the binomial coefficients)
We raise this because the path integral approach is sometimes employed in time travel problems, but turns out to be unwieldy in this simple problem. Expressions like (\ref{eq:S_frac_prop}) instead come from standard constructions that are well-known in feedback control and the ``renormalization'' of counting multiple passes around a loop already catered for in the denominator. In other words, just because you can do a sum-over-all-paths, it does not mean that you should!

\subsection{The Greenberger-Svozil Time Loop}
\begin{figure}[ht]
	\centering
		\includegraphics[width=0.50\textwidth]{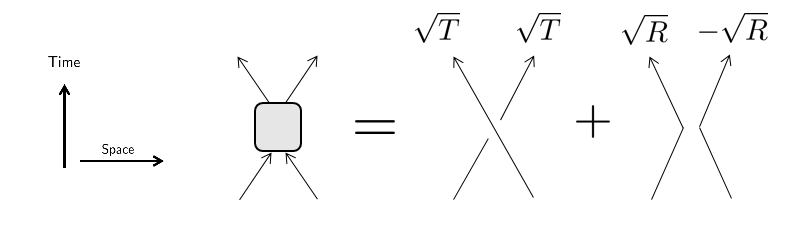}
	\caption{The beamsplitter matrix is given an explicit realization.}
	\label{fig:BS}
\end{figure}

In the time loop considered by Greenberger and Svozil \cite{GS05}, the beamsplitters
were both taken to be the same (we slightly modify the signs from their original matrix)
\begin{eqnarray}
S^A=S^B 
=\left[ 
\begin{array}{cc}
\sqrt{R} & \sqrt{T} \\ 
\sqrt{T} & - \sqrt{R}
\end{array}
\right] 
\label{eq:BS_matrix}
\end{eqnarray}
where $R,T>0$ are the reflection and transmission coefficients and satisfy $R+T=1$ (see Figure \ref{fig:BS}). 

We now find, 
\begin{eqnarray*}
S=\left[ 
\begin{array}{cc}
RG_1+TG_2 & \sqrt{RT}\left( G_1 -G_2 \right)  \\ 
-\sqrt{RT}\left( G_1 - G_2 \right)  & TG_{1}+RG_{2}
\end{array}
\right] ,
\end{eqnarray*}
from which we arrive at 
\begin{eqnarray}
S_{\text{fb}}=TG_{1}+RG_{2}+\frac{RT(G_{1}-G_{2})^{2}M}{1-(RG_1+TG_2 )M}.
\label{eq:TF_GS}
\end{eqnarray}

Alternatively, we compute the transfer coefficients and one readily find: 
$ \kappa_{\mathrm{I}} = T$, $\kappa_{\mathrm{II}} =R$, $\kappa_{\mathrm{III}} = -TR$, $\kappa_{\mathrm{IV}} = -TR$, $\kappa_{\mathrm{VII}} = TR$, $\kappa_{\mathrm{VIII}} = RT$ and $\beta_{\mathrm{I}} = R$, $\beta_{\mathrm{II}} = T$, $\beta_{\mathrm{V}} =T$, $\beta_{\mathrm{VI}} = R$.

In the limiting cases of perfect reflection and perfect transmission we find
(here we should look to the paths I and II, respectively, in Figure \ref{fig:GS_paths} and ignore the closed loops!) 
\begin{eqnarray}
\lim_{T\rightarrow 1}S_{\text{fb}}=G_{1},\quad \lim_{R\rightarrow 1}S_{\text{fb}}=G_{2} .
\end{eqnarray}

\begin{remark}
In \cite{GS05}, the matrix is $\begin{bmatrix}
\sqrt{R} & -i \sqrt{T} \\ 
-i\sqrt{T}  & \sqrt{R}
\end{bmatrix} $, but need to introduce an extra factor of -1 into the second branch of the interferometer which they refer to as a ``phasing effect''. Our choice is simpler and does not require additional features.
\end{remark}

\begin{remark}
\textbf{Reversing the Polarity of the Later Beamsplitter}
Let us replace the second beamsplitter with
\begin{eqnarray*}
S^{B}=\left[ 
\begin{array}{cc}
\sqrt{T} &  - \sqrt{R} \\ 
\sqrt{R} & \sqrt{T}
\end{array}
\right] .
\end{eqnarray*}
Note that the parameter $R$ is the reflection coefficient of the beamsplitter at $A$ while also the transmission coefficient at $B$. Either by recalculating the matrix $S$ from (\ref{eq:S_tot}) or by computing the new factors $\kappa_{\mathrm{I}}$, etc., we obtain
\begin{eqnarray*}
S_{\text{fb}}=  \sqrt{RT} (G_{1}-G_{2})-\frac{(RG_1 +TG_2 )(RG_{2}+TG_{1})M}{1-  \sqrt{RT} (G_1-G_2)M}.
\end{eqnarray*}
In the limit of perfect transmission ($T \to 1$), for instance, we obtain $S_{\text{fb}} = G_2 MG_1$, and this corresponds to path III in Figure \ref{fig:GS_paths}. $R \to 1$ would then be path IV.
\end{remark}

The welcher weg approach leads naturally to the path integral formalism: here we sum over all the probability amplitudes calculated from the propagators and beamsplitter coefficients associated with all possible paths allowing for multiple loops. So far we have been able to keep track of terms using sum rules due to the abelian nature of the problem (our Hilbert space is just $\mathbb{C}$ and everything is commutative). However, things are starting to get unwieldily and it is difficult to believe that a sum-over-all-paths approach will be useful. The more fruitful approach is the traditional control engineer's approach of using fractional linear transformation (M\"{o}bius transformations) with the sum over paths being more compactly handled: any path expansion really should be consider as arising from the multi-loop expansion (\ref{eq:multiloop_expansion}).

\section{Grandfather Paradoxes}
\label{sec:Grandfather}

\subsection{Death By Quantum Measurement}
\begin{figure}[ht]
	\centering
		\includegraphics[width=0.500\textwidth]{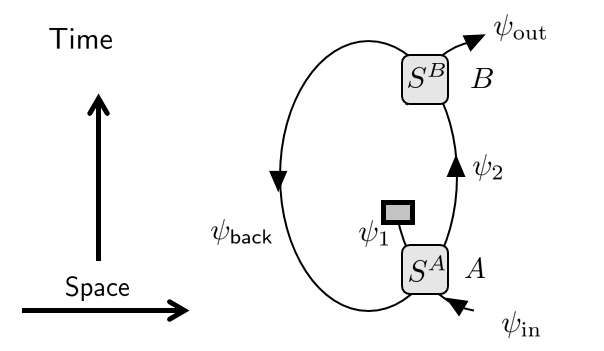}
	\caption{A detector is placed inside first branch of the interferometer (just outside the first output port). If it detects the particle then it absorbs it.}
	\label{fig:GS_loop_meas}
\end{figure}

The first method is to place an absorber in one of the branches of the interferometer, see Figure \ref{fig:GS_loop_meas}. It is natural to interpret this as a quantum measurement - specifically of demolition type because we destroy the particle if we detect it. 

There are only two possibilities here: either the particle comes out and is detected or it does not. Of course, if it is not detected then it must come out at $B$ from the external output port.
This time we must have
\begin{eqnarray}
    |\psi_{\text{in}} (t_A) |^2 =   |\psi_1 (t_A)  |^2 +
    |\psi_{\text{out}} (t_B) |^2 .
\end{eqnarray}

It is instructive to work through the calculation here. The input-output relations at $B$ change to 
\begin{eqnarray}
\left[ 
\begin{array}{c}
 \psi _{\text{back}} \\
\psi _{\text{out}} 
\end{array}
\right]_B =S^B \left[ 
\begin{array}{c}
0 \\ 
\psi _{2}
\end{array}
\right]_B ,
\end{eqnarray}
because $\psi_1$ no longer makes a contribution. The overall effect of this is that we now arrive at the equation (\ref{eq:pre-link}) but with a new transfer matrix $\tilde{S}$ given by
\begin{eqnarray}
\tilde{S} = S^B 
\left[ 
\begin{array}{cc}
0& 0 \\ 
0 & G_2
\end{array}
\right] 
S^A 
\equiv
\left[ 
\begin{array}{cc}
\tilde{S}_{\text{back,back}} & \tilde{S}_{\text{back},\text{in}} \\ 
\tilde{S}_{\text{out,}\text{back}} & \tilde{S}_{\text{out,in}}
\end{array}
\right]
.
\label{eq:S_tilde}
\end{eqnarray}
We may now eliminate the feedback path as before to now arrive at 
\begin{eqnarray}
\psi _{\text{out}}\left( t_{B}\right) = \tilde{S}_{\text{fb}}\,\psi _{\text{in}%
}\left( t_{A}\right)
\end{eqnarray}
where the new transfer coefficient is
\begin{eqnarray}
\tilde{S}_{\text{fb}} =
\text{M\"{o}b} ( \tilde{S} , M ).
\label{eq:S_fb_tilde}
\end{eqnarray}

\begin{proposition}
    The $\psi_{\text{in}} (t_A)$ to $\psi_{\text{out}} (t_B)$ transfer coefficient $\tilde{S}_{\text{fb}}$ for the set up in Figure \ref{fig:GS_loop_meas} is obtained from the original one $S_{\text{fb}}$ given in (\ref{eq:S_fb}) for the set up in Figure \ref{fig:GS_loop} by mathematically setting $G_1=0$.
\end{proposition}

This is indeed what Greenberger and Svozil do in \cite{GS05}, but it is worth checking through that this is the correct thing to do! It is routine now to see that
\begin{eqnarray*}
    \tilde{S}_{\text{fb}}&=& 
    S^B_{\text{out},2} G_2 S^A_{2, \text{in}} \nonumber \\
    && +
    S^B_{\text{out},2} G_2 S^A_{2, \text{back}}   M (1 - S^B_{\text{back},2} G_2 S^A_{2, \text{back}} M )^{-1} S^B_{\text{back},2} G_2 S^A_{2, \text{in}},
\end{eqnarray*}
and that $\psi_1 (t_A) = \tilde{S}_1 \, \psi_{\text{in}} (t_A)$ where
\begin{eqnarray*}
    \tilde{S}_1 &=& S^B_{1, \text{in}} \nonumber \\
    &&
    + S^A_{1, \text{back}}
    M  (1 - S^B_{\text{back},2} G_2 S^A_{2, \text{back}} M )^{-1} S^B_{\text{back},2} G_2 S^A_{2, \text{in} }.
\end{eqnarray*}

\subsubsection{Quantum Guardian Angel}
We may ask whether it is possible to ``tune'' the interferometer to ensure that the particle never gets detected at $A$.

This would mean that we arrange our model so that $\psi_1 (t_A) \equiv 0$ for arbitrary input $\psi_{\text{in}}$.

\begin{figure}[ht]
	\centering
		\includegraphics[width=0.2500\textwidth]{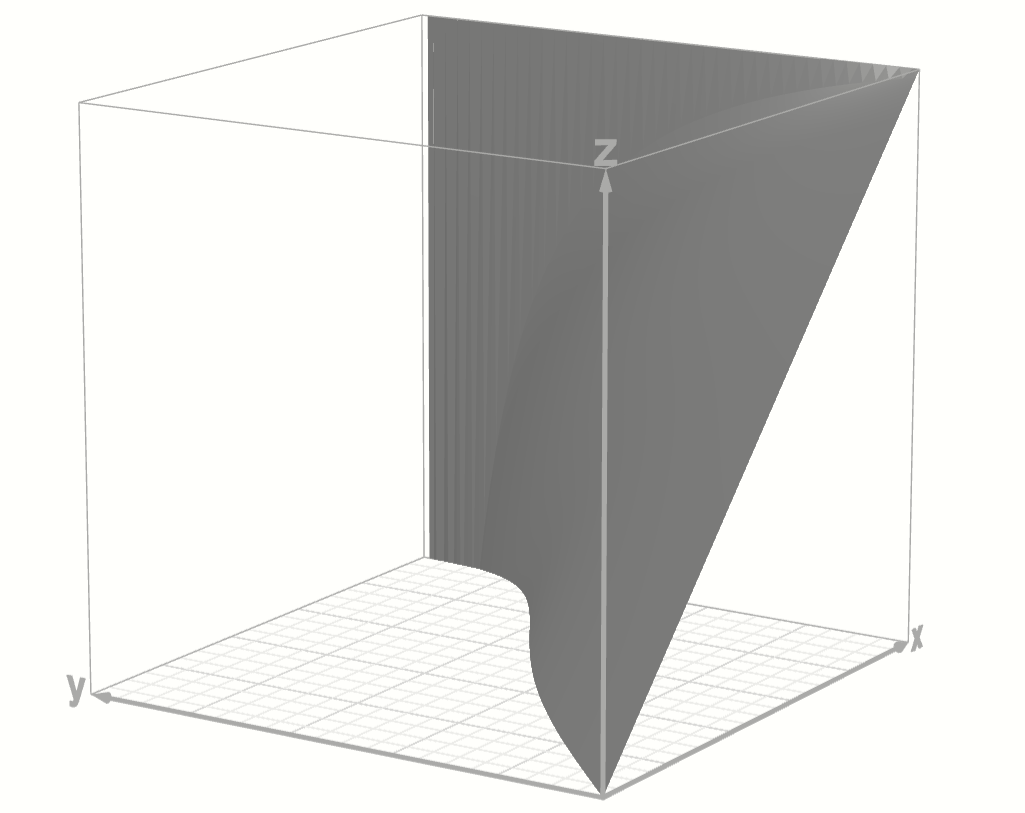}
	\caption{Plot of $z=f(x,y)$ in the region $0 < x,y < 1$ for $z\ge 0$.}
	\label{fig:parameters}
\end{figure}

Let us take the beamsplitter matrices $S^A$ and $S^B$ to have the same form as in (\ref{eq:BS_matrix}) but allow them to have different reflectivity parameters $R_A$ and $R_B$, respectively. Here we see that $\tilde{S}_1 = \tau_A -\sqrt{1-\tau_A^2}\frac{\tau_B}{1- \tau_AG\tau_B M} G_2$, where $\tau_A= \sqrt{T_A}$ and $\tau_B = \sqrt{T_B}$. For simplicity, take $M= G_1^{-1}$ which does says that the backward propagator undoes the forward one). We then obtain $ \tilde{S}_1 = f( \tau_A, \tau_B ) $ where
\begin{eqnarray}
        f(x,y )  =  x  -\sqrt{1- x^2}\frac{y}{1- xy}
\end{eqnarray}
The plot of $z= f(x,y)$ is given below (Figure \ref{fig:parameters} for $ 0 \ < x,y <1$ and $z \ge 0$. We note that for each $x$ in the range there will be (unique) $y$ in the range such that $z=0$, and \textit{vice versa}. It is therefore always possible to arrange $\tilde S_1 \equiv 0$ by adjusting the reflectivity parameter of either one of the beamsplitters.
    
\subsubsection{General Hilbert Space Formulation}
\begin{figure}[ht]
	\centering
		\includegraphics[width=0.50\textwidth]{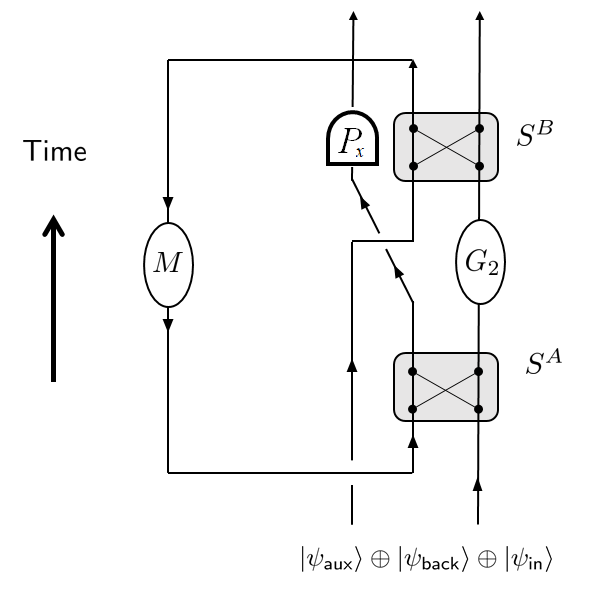}
	\caption{This is the generalization of Fig. \ref{fig:GS_loop_meas} to the multi-dimensional Hilbert space setting. Rather than a simple in-loop measurement to see if the particle is there, we instead consider a projective measurement $P$.}
	\label{fig:auxi}
\end{figure}

We now extend this to the general Hilbert space setting. Again we include the measurement of the state along the first branch of the interferometer then we have a slightly more involved construction to consider, see Figure \ref{fig:auxi}. 

This entails including a additional Hilbert space $\mathfrak{h}_{\text{aux}}$. Strictly speaking, this should have been present in our earlier treatment however was omitted because we just took the input auxiliary state to be $|\psi _{\text{aux}}\rangle =0$ - that is, assumed there was zero input to the beamsplitter at $B$. 

The in-loop measurement no longer has to be a complete demolition of the particle. We may take it to be the measurement of some observable $X = \sum_x xP_x$ where $P_x$ is the orthogonal projection onto the eigenspace of eigenvalue $x$. Therefore, if the measurement result is $x$ (this will be classical information that is then manifest to external observers), we should apply the orthogonal projection $P_x$, see Figure \ref{fig:auxi}. 

This time round, the open loop unitary on $\mathfrak{h}_{\text{aux}}\oplus \mathfrak{h}_{\text{back}}\oplus \mathfrak{h}_{\text{in}}$ takes the block form
\begin{eqnarray*}
S=\left[ 
\begin{array}{cc}
P & 0\,0 \\ 
\begin{array}{c}
0 \\ 
0
\end{array}
& S^{B}
\end{array}
\right] \left[ 
\begin{array}{ccc}
0 & 1 & 0 \\ 
1 & 0 & 0 \\ 
0 & 0 & G_{2}
\end{array}
\right] \left[ 
\begin{array}{cc}
1 & 0\,0 \\ 
\begin{array}{c}
0 \\ 
0
\end{array}
& S^{A}
\end{array}
\right] .
\end{eqnarray*}

We now identify the external space as $\mathfrak{h}_{\text{ext}}\cong \mathfrak{h}_{\text{aux}}\oplus \mathfrak{h}_{\text{in}}$ and the internal feedback space as $\mathfrak{h}_{\text{int}} =\mathfrak{h}_{\text{back}}$. The M\"{o}bius transformation can be performed as before, except the external space is split into two orthogonal components. However, $S$ is only a contraction this time due to the presence of $P_x$.  Likewise, $S_{\text{fb}}$ will be a contraction on $\mathfrak{h}_{\text{ext}}\cong \mathfrak{h}_{\text{aux}}\oplus
\mathfrak{h}_{\text{in}}$ and can be decomposed as

\begin{eqnarray*}
S_{\text{fb}}\equiv \left[ 
\begin{array}{cc}
\left[ S_{\text{fb}}\right] _{\text{aux,aux}} & \left[ S_{\text{fb}}\right]
_{\text{aux,in}} \\ 
\left[ S_{\text{fb}}\right] _{\text{out,aux}} & \left[ S_{\text{fb}}\right]
_{\text{out,in}}
\end{array}
\right] 
\end{eqnarray*}
Taking the initial external state as $0\oplus |\psi _{\text{in}}\rangle $ (that is, we have zero auxiliary input), we get output $|\psi _{\text{aux}}^{\prime }\rangle \oplus |\psi _{\text{out} }\rangle $ where

\begin{eqnarray*}
|\psi _{\text{aux}}^{\prime }\rangle  &=&\left[ S_{\text{fb}}\right] _{\text{%
aux,in}}|\psi _{\text{in}}\rangle , \\
|\psi _{\text{out}}\rangle  &=&\left[ S_{\text{fb}}\right] _{\text{out,in}%
}|\psi _{\text{in}}\rangle .
\end{eqnarray*}
The operator $\left[ S_{\text{fb}}\right] _{\text{out,in}}$ now corresponds
to what we previously denoted as $\tilde{S}_{\text{fb}}$. 

\subsection{Death By Time-Loop}
The second method for dispatching your grandfather is to arrange things so that the open-loop matrix $S$ is block diagonal:
\begin{eqnarray}
    S \equiv
    \left[ 
\begin{array}{cc}
G_1 & 0 \\ 
0 & G_2
\end{array}
\right].
\label{eq:diagonal}
\end{eqnarray}
In \cite{GS05}, this is achieved by taking $G_1$ and $G_2$ to be the same (let us call the common propagator $G$) and using the fact that $S^A$ and $S^B$ are inverses to each other in their case
\begin{eqnarray}
S= S^B 
\left[ 
\begin{array}{cc}
G & 0 \\ 
0 & G
\end{array}
\right] 
S^A  \equiv 
\left[ 
\begin{array}{cc}
G & 0 \\ 
0 & G
\end{array}
\right].
\end{eqnarray}

Effectively, the Mach-Zehnder interferometer is now balanced: particle 1 always outputs from the first port and particle 2 always outputs from the second. We obtain the decoupled equations $\psi_{\text{out}} (t_B) = G_2 \, \psi_{\text{in}} (t_A)$ and $\psi_{\text{back} } (t_B) = G_1 \psi_{\text{back}} (t_A)$. We also have $\psi_{\text{back}} (t_A) = M \, \psi_{\text{back}} (t_B )$, and it follows that we should take $M \equiv G_1^{-1}$ for consistency.

\section{Time Traveling Harmonic Oscillators}
While the general approach to quantum information processing is displayed in terms of qubits (or qudits), in practice, a more natural way to encode quantum information is in terms of continuous observables, and here quantized harmonic oscillators offer the most suitable framework \cite{Furusawa_Loock}.

We now consider a tractable example of out approach where the system consists of $n$ harmonic oscillators. In this case, $\mathfrak{h}$ is infinite-dimensional. Let $\boldsymbol{\alpha} = (\alpha_1 , \cdots , \alpha_n ) \in \mathbb{C}^n$, and denote the coherent state with complex amplitudes $\boldsymbol{\alpha}$ by $| \boldsymbol{\alpha} \rangle$. That is, if $b_k$ is the annihilator of the $k$th oscillator, then $ (b_k - \alpha_k )| \boldsymbol{\alpha} \rangle =0$.

\begin{figure}[ht]
	\centering
		\includegraphics[width=0.30\textwidth]{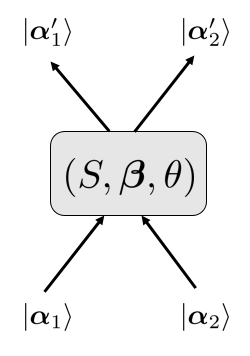}
	\caption{A linear device $(S, \boldsymbol{\beta}, \theta )$ transforming incoming coherent states $ | \boldsymbol{\alpha}_1 \oplus \boldsymbol{\alpha}_2 \rangle$ to outgoing states  $e^{i\theta} |\boldsymbol{\alpha}_1^\prime \oplus \boldsymbol{\alpha}_2^\prime \rangle$ according to (\ref{eq:linear_trans}). }
	\label{fig:BS_coherent}
\end{figure}

In place of beamsplitters, we more generally consider devices that perform unitary scattering ($S$), a Weyl displacement $(\boldsymbol{\beta})$, and a phase change $(e^{i\theta })$. This is a linear device which we denote as $(S, \boldsymbol{\beta}, \theta )$, see Figure \ref{fig:BS_coherent}.

We define the Weyl unitaries by
\begin{eqnarray}
    W(S,\boldsymbol{\beta}, \theta ) | \boldsymbol{\alpha} \rangle
    =e^{i \theta} | S\boldsymbol{\alpha} + \boldsymbol{\beta} \rangle
    .
\end{eqnarray}
Here $\theta $ is a real phase, $\boldsymbol{\beta} \in \mathbb{C}^n$ is a displacement, and $S$ is a unitary on $\mathbb{C}^n$.

Note that $W(S^B,\boldsymbol{\beta}^B, \theta^B )W(S^A,\boldsymbol{\beta}^A, \theta^A )$ is again a unitary with parameters now given by
\begin{gather}
    (S^B,\boldsymbol{\beta}^B, \theta^B ) \triangleleft (S^A,\boldsymbol{\beta}^A, \theta^A )=\nonumber \\ (S^BS^A,\boldsymbol{\beta}^B +S^B \boldsymbol{\beta}^A, \theta^A+\theta^B +
    \mathrm{Im} \boldsymbol{\beta}^{B \ast}S^B \boldsymbol{\beta}^A).
\end{gather}
This product is the group law for the extended Heisenberg group where we work with annihilators rather than quadratures: the additional phase picked up is due to the Weyl commutation relations.

We now consider generalizing our beamsplitters so that they act on incoming coherent states $ | \boldsymbol{\alpha}_1 \oplus \boldsymbol{\alpha}_2 \rangle$ to produce an output $e^{i\theta} |\boldsymbol{\alpha}_1^\prime \oplus \boldsymbol{\alpha}_2^\prime \rangle$ where
\begin{eqnarray}
|\boldsymbol{\alpha}_1^\prime \rangle &=&
S_{11} |\boldsymbol{\alpha}_1 \rangle
+S_{12} |\boldsymbol{\alpha}_2 \rangle
+|\boldsymbol{\beta}_1 \rangle , \\
|\boldsymbol{\alpha}_2^\prime \rangle &=&
S_{21} |\boldsymbol{\alpha}_1 \rangle
+S_{22} |\boldsymbol{\alpha}_2 \rangle
+|\boldsymbol{\beta}_2 \rangle
.
\label{eq:linear_trans}
\end{eqnarray}
Here, $(S, \boldsymbol{\beta}, \theta )$ includes the unitary $S=\begin{bmatrix}
    S_{11} & S_{12} \\
    S_{21} & S_{22}
\end{bmatrix}$ and the displacement $\boldsymbol{\beta}=\begin{bmatrix}
    \boldsymbol{\beta}_1 \\
    \boldsymbol{\beta}_2
\end{bmatrix}$, see Figure \ref{fig:BS_coherent}.

We may combine $(S^A, \boldsymbol{\beta}^A, \theta^A )$ at time $t_A$ with $(S^B, \boldsymbol{\beta}^B, \theta^B )$ to get $(S, \boldsymbol{\beta}, \theta )$ given by
\begin{eqnarray}
    (S, \boldsymbol{\beta}, \theta ) &=&
    (S^B, \boldsymbol{\beta}^B, \theta^B ) \triangleleft (G_{BA}, 0 , 0) 
    \triangleleft (S^A, \boldsymbol{\beta}^A, \theta^ A)\nonumber \\
    &=&
    (S^BG_{BA}S^A, \boldsymbol{\beta}^B +S^BG_{BA} \boldsymbol{\beta}^A, \nonumber\\
    &&\theta^A + \theta^B + \mathrm{Im} \boldsymbol{\beta}^{B \ast}S^B G_{BA}\boldsymbol{\beta}^A). 
\end{eqnarray}
Here $G_{BA} = \begin{bmatrix}
    G_1 &0 \\
    0  & G_2 
\end{bmatrix}$.

One then introduces the model matrix
\begin{eqnarray}
    \mathsf{V} (S, \boldsymbol{\beta}, \theta )
    = \begin{bmatrix}
        -\frac{1}{2 } \boldsymbol{\beta}^\ast \boldsymbol{\beta} -i \theta & - \boldsymbol{\beta}^\ast S \\
        \boldsymbol{\beta} & S
    \end{bmatrix}
    .
\end{eqnarray}

The feedback time-loop arises from setting $|\boldsymbol{\alpha}_1 \rangle = M\, | \boldsymbol{\alpha}_1^\prime \rangle$ and this leads to the reduced model matrix
\begin{eqnarray}
    \mathsf{V}_{\mathrm{fb}} = \text{M\"{o}b}_2 \bigg( \mathsf{V} (S, \boldsymbol{\beta}, \theta ) , M \bigg) .
\end{eqnarray}
We note that $\mathsf{V}_{\mathrm{fb}}$ takes the form $\mathsf{V} (S_{\mathrm{fb}}, \boldsymbol{\beta}_{\mathrm{fb}} , \theta_{\mathrm{fb}})$ where $S_{\mathrm{fb}}= \text{M\"{o}b}_2 \big( S , M \big)=S_{22}+ S_{21} M (1-S_{11} M)^{-1} S_{21} $ as before, while
\begin{eqnarray}
    \boldsymbol{\beta}_{\mathrm{fb}} = \beta_2 + S_{21} M (1- S_{11}M)^{-1}
    \boldsymbol{\beta}_1 .
\end{eqnarray}

\section{Continuous Signals}
Up until now we have considered sending a quantum mechanical system back in time. More generally we could consider a send a quantum field. Here we use the language of quantum input-output theory \cite{GC}. Our input will be a bosonic white noise input $b_{\mathrm{in},j} (t)$ satisfying singular commutation relations $[b_{\mathrm{in},j} (t) , b_{\mathrm{in}, k} (s) ] = \delta_{jk} \delta (t-s)$.

\begin{figure}[ht]
	\centering
		\includegraphics[width=0.60\textwidth]{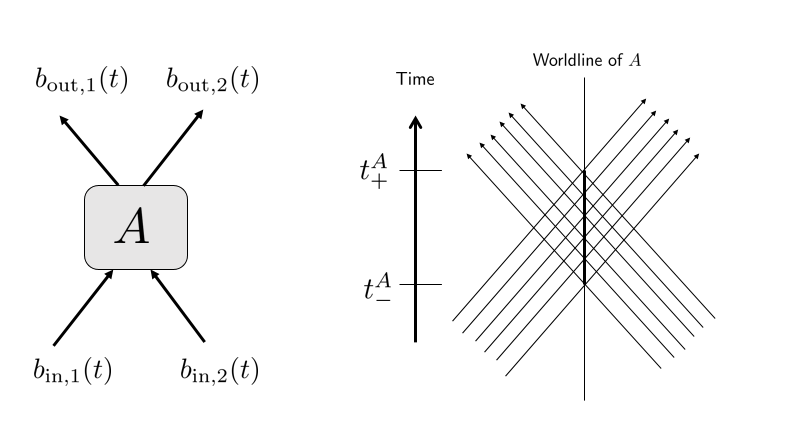}
	\caption{(left) system $A$ is driven by inputs $b_{\mathrm{in}, j} $ for $t_-^A < t< t^A_+$; (right) the spacetime view of the input-outputs.}
	\label{fig:BS_time}
\end{figure}

At the first device $A$ we can consider a quantum system with its own Hilbert space $\mathfrak{h}_A$, see Figure \ref{fig:BS_time} (left). This time we may specify a triple $(S^A =\begin{bmatrix}
    S_{11}^A & S_{12}^A \\
    S_{21}^A & S_{22}^A
\end{bmatrix}, L^A =\begin{bmatrix}
    L^A_{1} \\
    L^A_{2}
\end{bmatrix}, H^A)$ where this time the entries $S^A_{jk}, L_j, H$ are operators and are not assumed to commute amongst themselves.
Overall, $S$ is still required to unitary and $H$ to  be self-adjoint.

The quantum signal is to be transmitted over a time window: $t^A_-$ being the start of transmission and $t^A_+$ being the end. See Figure \ref{fig:BS_time} (right).

The input-output relations are given by
\begin{eqnarray}
    b_{\mathrm{out},j} (t)= S_{jk }^A(t) \, b_{\mathrm{in},k} (t) +L_j^A (t) .
\end{eqnarray}
Note that the coefficients are understood as being in the Heisenberg picture under the open systems dynamics arising from $H^A$ and the inputs. In general, The dynamics of the device itself is described by (we suppress the time-dependence of the system operators for convenience)
\begin{eqnarray}
    \dot X &=& b_{\mathrm{in},j} (t)^\ast ( S_{lj}^\ast XS_{lk} -X ) b_{\mathrm{in},k} (t)
    \nonumber \\
    &+& b_{\mathrm{in},k}^\ast S_{jk} ^\ast  [X,L_k]  (t) 
    +  [L_j^\ast , X] S_{jk} b_{\mathrm{in},k}  \nonumber\\
    &+& \frac{1}{2} L^\ast_k [X,L_k]+\frac{1}{2} [L^\ast_k , X] L_k -i[X,H] .
\end{eqnarray}
On its own, each component is described by a quantum Markovian model. 
The full quantum model for this has been presented in \cite{SLH} where the quantum field signals propagated along the edges of the network by simple time-shifts. In the present case, the edges are replaced by paths and we may allow for more general propagation.

\section{Conclusion}
Our aim here was to adapt standard methods for studying quantum feedback networks to address foundational questions involving quantum physics and time travel. The main mathematical tool which emerges is the M\"{o}bius transformation. The application is similar to the quantum feedback network theory insofar as we are handling quantum evolutions with feedback and need to be careful of the noncommutative nature of the objects under consideration. The general theory turns out to be reasonably tractable, but surprisingly rich and we are able to extend beyond the complex probability amplitudes considered originally in \cite{GS05}.

The formalism presented allows us to consider a time-traveling quantum system with a general Hilbert space $\mathfrak{h}_0$. We treated explicitly the finite dimensional case $\mathbb{C}^d$ where the non-abelian nature shows that the M\"{o}bius transformations efficiently keep track of the multiple loop behaviour, which would be extremely cumbersome in a path-integral analysis: plus it allows for component devices (generalizing beamsplitters) with their own quantum degrees of freedom which may couple with the time-traveling system. We also showed that linear quantum systems, which have an infinite dimensional Hilbert space, may also be treated in a tractable manner. 

Up until now, the usual picture is of particle instantaneously entering a time machine however a more natural description would be to allow for a continuous quantum field to enter over a time interval. This was considered in the final section. This opens up the possibility, at  least, of performing continuous monitoring of the eventual output signals and potentially controlling the past.

\section*{Acknowledgments}
The author is grateful to Joshua Combes for introducing him to references \cite{Pegg01} and \cite{GS05} and pointing out their potential link to quantum feedback networks.


\end{document}